\documentclass[twocolumn,noshowpacs,noshowkeys,pra,aps,superscriptaddress,bibliography]{revtex4-1}%
\usepackage{graphicx}
\usepackage{color,pgfplots}
\usepackage{epstopdf}
\usepackage{braket}
\usepackage{amsmath}
\usepackage{notes2bib}
\usepackage{pdfpages}

\newcommand{\Fig}[1]{Figure \ref{#1}}   
\newcommand{\fig}[1]{fig.\ref{#1}}   

\newcommand{\highlight}{} 

\begin{document}

\title{The stability of polarisation singularities in disordered photonic crystal waveguides}

\author{Ben Lang}
\email[Contact me at: ]{bl9453@bristol.ac.uk}
\affiliation{Department of Electrical and Electronic Engineering, University of Bristol, Merchant Venturers Building, Woodland Road, Bristol, BS8 1UB, UK}

\author{Daryl M. Beggs}
\affiliation{Centre for Quantum Photonics, H.H. Wills Physics Laboratory, University of Bristol, Tyndall Avenue, Bristol, BS8 1TL, UK}

\author{Andrew B. Young}

\author{John G. Rarity}
\affiliation{Department of Electrical and Electronic Engineering, University of Bristol, Merchant Venturers Building, Woodland Road, Bristol, BS8 1UB, UK}

\author{Ruth Oulton}
\affiliation{Department of Electrical and Electronic Engineering, University of Bristol, Merchant Venturers Building, Woodland Road, Bristol, BS8 1UB, UK}
\affiliation{Centre for Quantum Photonics, H.H. Wills Physics Laboratory, University of Bristol, Tyndall Avenue, Bristol, BS8 1TL, UK}

\begin{abstract}
The effects of short range disorder on the polarisation characteristics of light in photonic crystal waveguides were investigated using finite difference time domain simulations with a view to investigating the stability of polarisation singularities. It was found that points of local circular polarisation (C-points) and contours of linear polarisation (L-lines) continued to appear even in the presence of high levels of disorder, and that they remained close to their positions in the ordered crystal. These results are a promising indication that devices exploiting polarisation in these structures are viable given current fabrication standards.
\end{abstract}


\maketitle

\section{Introduction}

A polarisation singularity occurs at a position in a vector field where one of the parameters describing the local polarisation ellipse (handedness, eccentricity or orientation) becomes singular \cite{Nye_cpoints}.  With the vector nature of electromagnetic fields, optics is an obvious place for the study and application of polarisation singularities, and they can be found in systems ranging from tightly focused beams \cite{Schoonover2006} to speckle fields \cite{Egorov2008,Flossmann2008} to photonic crystals \cite{c-points} and others \cite{Flossmann2005,Berry2004,transverse_long}.  For example, at circular polarisation points, the orientation of the polarisation ellipse is singular, whereas along contours of linear polarisation the handedness is singular.  The former are called C-points and the latter L-lines.  Recently, the use of polarisation singularities for quantum information applications is generating much interest, as they can couple the spin-states of electrons confined to quantum dots to the optical modes of a waveguide \cite{artur_andrew}.  For example, at a circular point (C-point), the sign of the local helicity/polarisation is governed by the propagation direction of the optical mode, which allows for spin-photon coupling to one direction only, sometimes referred to as the "chiral waveguide" effect \cite{sheffield,Rauschenbeutel,atom_fibre,atom_switch,rodriguez} \bibnote{Note: most "chiral waveguide" systems have positions in the electric field where the polarisation ellipse approaches circular with typical ellipticity reaching 0.92. In contrast photonic crystal waveguides display circular points (with ellipticity of $\pm 1$.)}.

A photonic crystal (PhC) is a periodic modulation of the refractive index, typically provided by etching holes in a transparent membrane, such as a semiconductor above its band-edge. PhC waveguides (PhCWGs) are strong candidates for finding applications for polarisation singularities.  As Bloch waves, the eigenmodes of PhCWGs possess a strong longitudinal, as well as transverse, component of their electric field profile.  The spatial dependence of both these components and the phase between them ensures a rich and complex polarisation landscape in the vicinity of the waveguide, and leads to the occurrence of polarisation singularities \cite{c-points,snom_excitation,immo_2015}.  However, any real system will inevitably contain imperfections that deviate away from this perfectly periodic pattern --- for example, the hole size and position can vary, or surface roughness caused by the etching. To be useful in applications, the polarisation structure should be robust to such real-world disorder.  However, to date no studies of how disorder affects the complex polarisation structure and existence of polarisation singularities has been conducted.

\begin{figure}[h]
\includegraphics[scale=0.25]{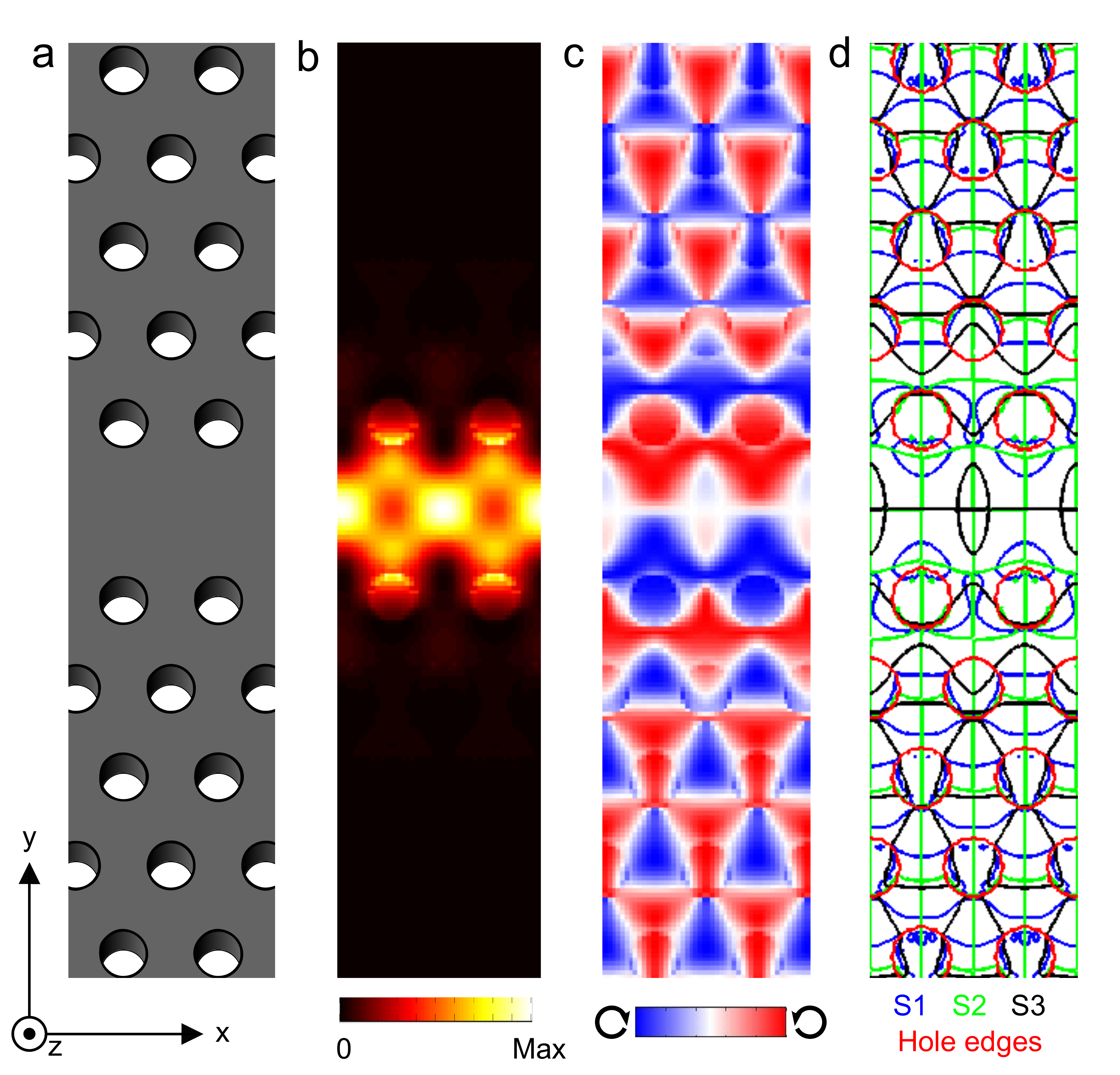}
\caption{Polarisation of eigenmode of ideal waveguide. (a) Schematic of PhCWG, showing gallium arsenide membrane ($n=3.54$) etched with air holes. (b) Eigenmode of ideal waveguide $|E|^2$.  (c) $S_3$, the extent to which the light is circularly polarised. (d) Zero-contours of the Stokes parameters $S_{1,2,3}$ shown by lines of the indicated shades. This figure, and all others in this paper, shows the fields at z=0 (centre of the slab).}
\label{mode}
\end{figure}

In this work, we use calculations of the eigenmodes of disordered PhCWGs to demonstrate that the polarisation singularities (C-points and L-lines) present in the ideal waveguide persist far beyond realistically expected levels of disorder.  By calculating the positions of these C-points and L-lines for a statistical ensemble of disordered waveguides, we demonstrate that not only do they persist, but their mean positions are close to those predicted for the ideal structure.  

\section{Methods}

Our PhCWG (shown in \fig{mode}\,(a)) is a single row of missing air holes from an hexagonal lattice of holes spaced $a$ apart and with a diameter of $d=0.48a$ in a dielectric slab with refractive index $n=3.54$  and thickness $h=0.56a$. The lattice constant $a$ can be chosen freely for different operating regimes, but with $a=250$\,nm  these values are suitable for a PhCWG in a GaAs membrane embedded with quantum dots emitting at 900\,nm.  We choose to work with a mode located at $ka/2\pi = 0.35$  with a eigenfrequency of $\omega a/2\pi c = 0.255$ {\highlight and group velocity $v_g=c/7$}. {\highlight We chose this mode as it possesses a significant number of C-points in the vicinity of the waveguide core, making it well suited to collect statistics when disorder is introduced.  Also the relatively modest slow-down factor ensures that the disorder localisation length remains longer than our waveguide supercell length, even for the largest disorder considered, ensuring that we remain in the ballistic photon transport regime and avoid complications caused by the formation of localised states in the disordered waveguides \cite{Mazoyer}.} The eigenmodes are calculated using a finite-difference time-domain (FDTD) method \cite{meep} with a grid resolution of $a/24$ and periodic boundary conditions in the direction of the waveguide (absorbing boundaries on all other edges).  In the course of the simulation a large number of modes are excited by dipole sources placed in the waveguide, but letting the simulation evolve in time causes most of these modes to decay such that the remaining power is in the waveguide's eigenmode $\mathbf{E}(\mathbf{r})$, as shown in \fig{mode}\,(b).

After finding the eigenmodes $\mathbf{E}(\mathbf{r})$, the polarisation structure in the vicinity of the waveguide is examined by calculating the four Stokes parameters, $S_{0,1,2,3}$, as a function of position $\mathbf{r}$. Together, the Stokes parameters completely specify the polarisation of the electric field on the Poincar\'{e} sphere \cite{textbook}:

\begin{equation}
        \begin{aligned}
	S_0 (\mathbf{r}) = |E_x (\mathbf{r})|^2 + |E_y (\mathbf{r})|^2, \\
	S_1 (\mathbf{r}) = (|E_x (\mathbf{r})|^2 - |E_y (\mathbf{r})|^2) / S_0 (\mathbf{r}), \\
	S_2 (\mathbf{r}) = 2*Re(E_x^* (\mathbf{r}) E_y (\mathbf{r}))/S_0 (\mathbf{r}),\\
	S_3 (\mathbf{r}) = 2*Im(E_x^* (\mathbf{r}) E_y (\mathbf{r}))/S_0 (\mathbf{r}).
	\end{aligned}
	\label{eqn2.qo}
\end{equation}

\begin{figure*}[t]
\includegraphics[scale=0.45]{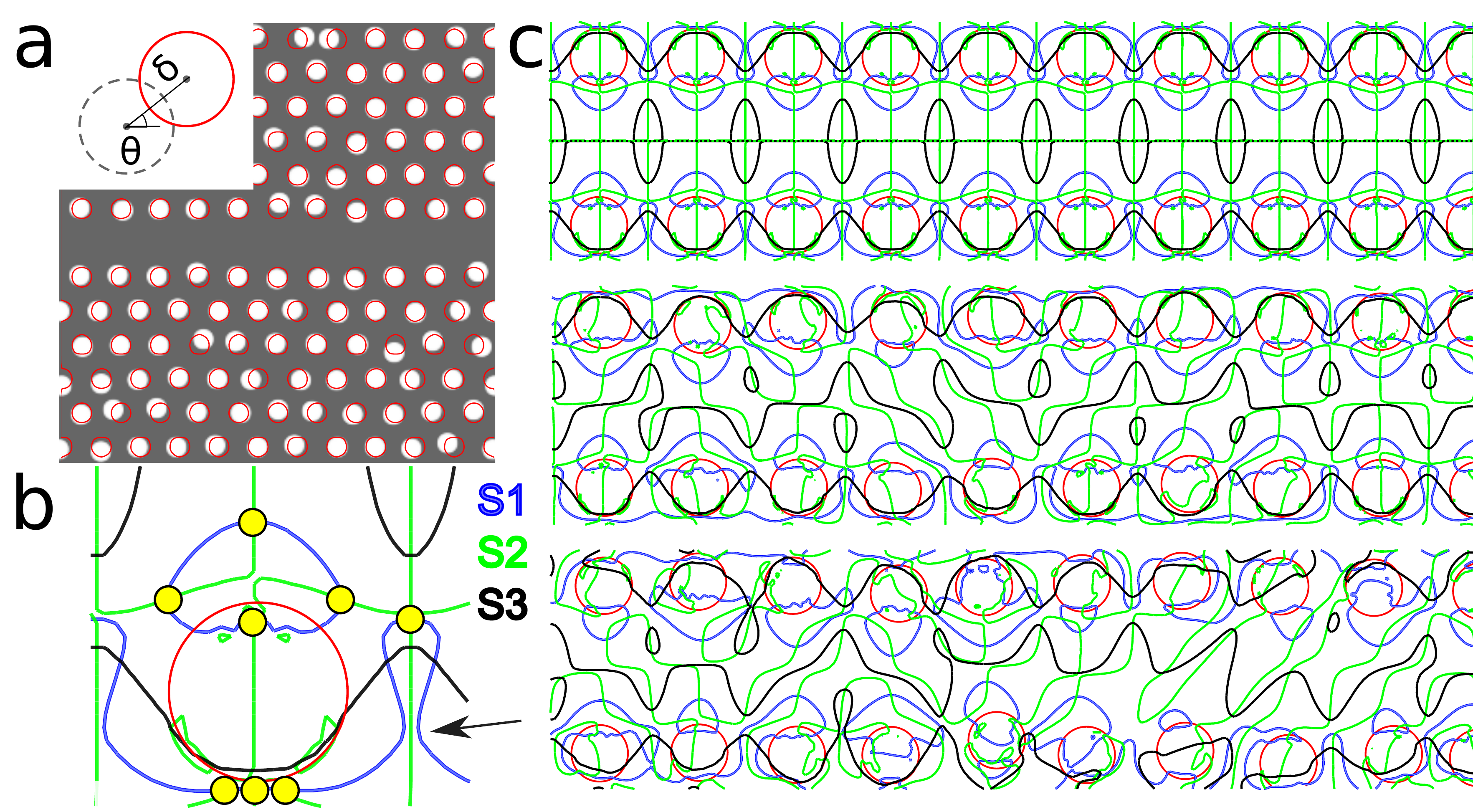}
\caption{Effect of disorder on Stokes parameters.  (a) Schematic of the disorder, the positions of the air holes in the ideal waveguide are shown in with lines, with a specific instance of a disordered crystal shaded in grey. Inset: the disorder model for an individual hole. (b) The zero-contours of $S_{1,2,3}$ are shown in blue, green and black respectively. The large circle denotes an air hole. C-points are marked with circles and L-lines by black lines. (c) More zero-contours (same colours), this time for supercells in three specific simulation runs with $\delta=$ 0, 0.03 and 0.09.}
\label{disorder}
\end{figure*}

$S_0$ specifies the total electric field strength, $|\mathbf{E}|^2$, and $-1\le S_{1,2,3} \le 1$ are the positions on the Poincar\'{e} sphere. \Fig{mode}\,(b) shows the calculated value of $S_0$ in the central plane ($z=0$) of the membrane for the ideal photonic crystal waveguide, and \Fig{mode}\,(c) shows the value of $S_3$.  $S_3$ is of particular interest to us, as it measures the extent of circular polarisation in the electric field profile.  Positions where $S_3=\pm 1$ are C-points with fully left and right circularly polarised fields, whereas contours where $S_3=0$ are L-lines with linear polarisation.  $S_1$ and $S_2$ indicate the degree of linear polarisation in the rectilinear and diagonal basis; specifically the values of $\pm 1$ in $S_1$ denotes vertical and horizontal polarization, while in $S_2$ they denote diagonal and anti-diagonal polarisation.  \Fig{mode}\,(d) shows the contours of $S_{1,2,3}=0$ for our PhCWG, with blue, green and black lines respectively.  The $S_3 = 0$ contour marks the L-lines, whereas the C-points are marked by the crossing of the $S_1 =0$ and $S_2 = 0$ contours (so points where $S_1 = S_2 =0$) , which coincide with positions where $S_3 = \pm 1$ (see \Fig{disorder}\,(b)) \cite{crossing_lines}.

\section{Disorder Results}

We now consider the effect of disorder on the Stokes parameters.  All disorder and imperfections in the waveguide are modelled as a random shift in the positions of the holes, with each hole displaced a small distance selected from a normal distribution with a mean of zero and a standard deviation of $a\delta$ in a random direction $\theta$ (selected from a uniform distribution $0 \le \theta < \pi$). The parameter $\delta$ therefore characterises the level of disorder in the simulation.  Although this model may not realistically represent all of the errors that occur in a typical fabrication process (it ignores the ellipticity, hole-size variation, side-wall roughness and angle), the displacement of the holes can be exaggerated in order to crudely account for these effects, and similar models of disorder are widely used \cite{high_q_disorder,loss_paper,same_disorder_model,korean,dmb2005}. Each simulation of a disordered waveguide used a supercell of dimensions $15a$ x $24a$ x $5a$, containing 15 unit-cells along the direction of the waveguide.  This length of 15 unit cells was chosen by checking that it is longer than the distance over which individual cells are significantly correlated with their neighbours {\highlight whilst being significantly shorter than the disorder localisation length (see discussion in supplementary information)}. An example part of disordered supercell is illustrated in \Fig{disorder}\,(a).

As described above, the position of the C-points are found in the eigenmodes of the disordered waveguides by examining where the zero contours of $S_1$ and $S_2$ cross, which is equivalent to finding positions where $S_3=\pm 1$.  \Fig{disorder}\,(b) shows the position of the C-points in one half of the unit cell of the ideal structure, in the vicinity of the hole closest to the waveguide.  The contours of $S_{1,2,3}=0$ are shown by blue, green and black lines respectively, and the C-points are shown by the yellow dots.  The L-lines are located simply by the contours where $S_3=0$ (black lines).  \Fig{disorder}\,(c) shows the response of the polarisation landscape to the introduction of disorder in the waveguide.  The top panel shows the ideal case ($\delta =0$), the middle shows $\delta=0.03$ and the bottom $\delta=0.09$.

\begin{figure}[t]
\includegraphics[scale=0.1]{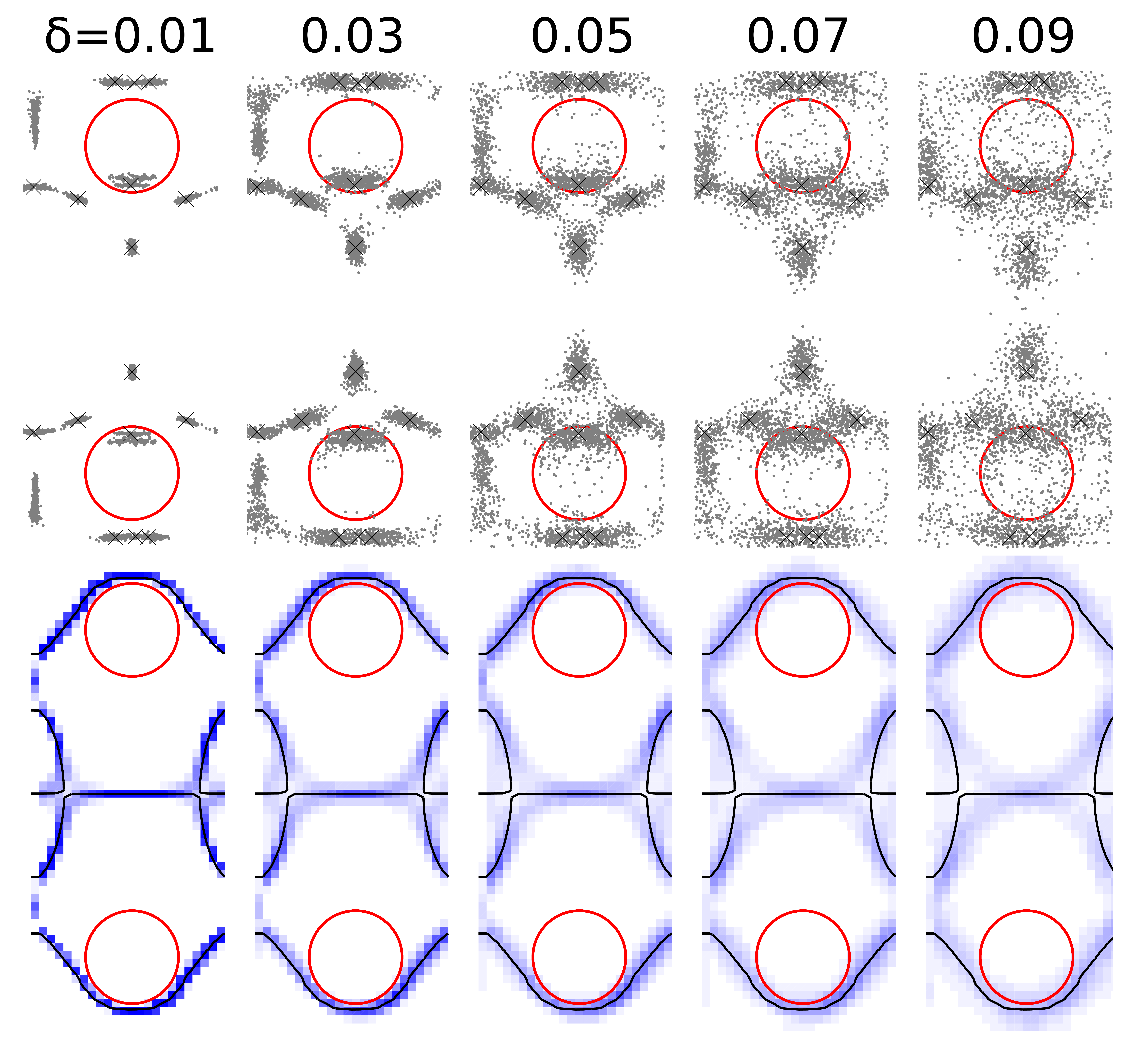}
\caption{Top: the locations of C-points. Each panel shows all the C-points found for a particular choice of $\delta$ folded into a single unit cell (using data from 32 separate simulations). Crosses: locations of the C-points with no disorder. Bottom: probability density of L-lines. Black lines: locations of the L-lines with no disorder. The circles (in both sets of panels) show the positions of the air holes (without disorder).}
\label{scatter}
\end{figure}

The most obvious feature apparent in \fig{disorder}\,(c) is that the zero-contours of the the Stokes parameters $S_{1,2,3}$ lose all sense of the symmetries present in the ideal structure.  In particular, the L-lines ($S_3 = 0$) lose the crossings seen in the core of the ideal waveguide, even for small disorder.  However, despite the apparent lack of order in \fig{disorder}\,(c), a more through analysis reveals that the polarisation singularities are remarkably robust to disorder.

Although not shown by \fig{disorder}\,(c) the number of points where the zero-contours of each Stokes parameter crosses itself falls rapidly with the introduction of disorder. Self crossings of this type imply the existence of two regions in which the Stokes parameter takes positive values and two where it takes negative ones meeting at a corner in a ``chess board'' like arrangement. In principle even the smallest amount of disorder will be sufficient to disturb this arrangement, creating a bridge connecting either the positive regions or the negative ones \cite{sign_rule,Nye_cpoints}. We find that these points disappear extremely rapidly with the introduction of disorder, and believe that the surviving cases can be attributed to the finite resolution of the simulations.

\Fig{scatter} shows a summary of the main findings of this work.  The top row considers the C-points.  Each panel in the top row marks with a blue dot the calculated positions of C-points found in each unit cell after 32 calculations performed with an independent realisation of the disordered structure for the values of the disorder parameter $\delta = $ 0.01, 0.03, 0.05, 0.07 and 0.09.  The black crosses mark the positions of the C-points in the ideal structure.  The bottom row considers the L-lines.  Each panel in the bottom row shows the location of the L-lines in the same disordered waveguides.  Specifically the probability of finding an L-line crossing a pixel in the calculation grid is plotted using a colour scale of white (zero probability) to dark blue (high probability) is shown. The L-lines in the ideal waveguide are shown by black lines.

\begin{figure}[h!]
\includegraphics[scale=0.35]{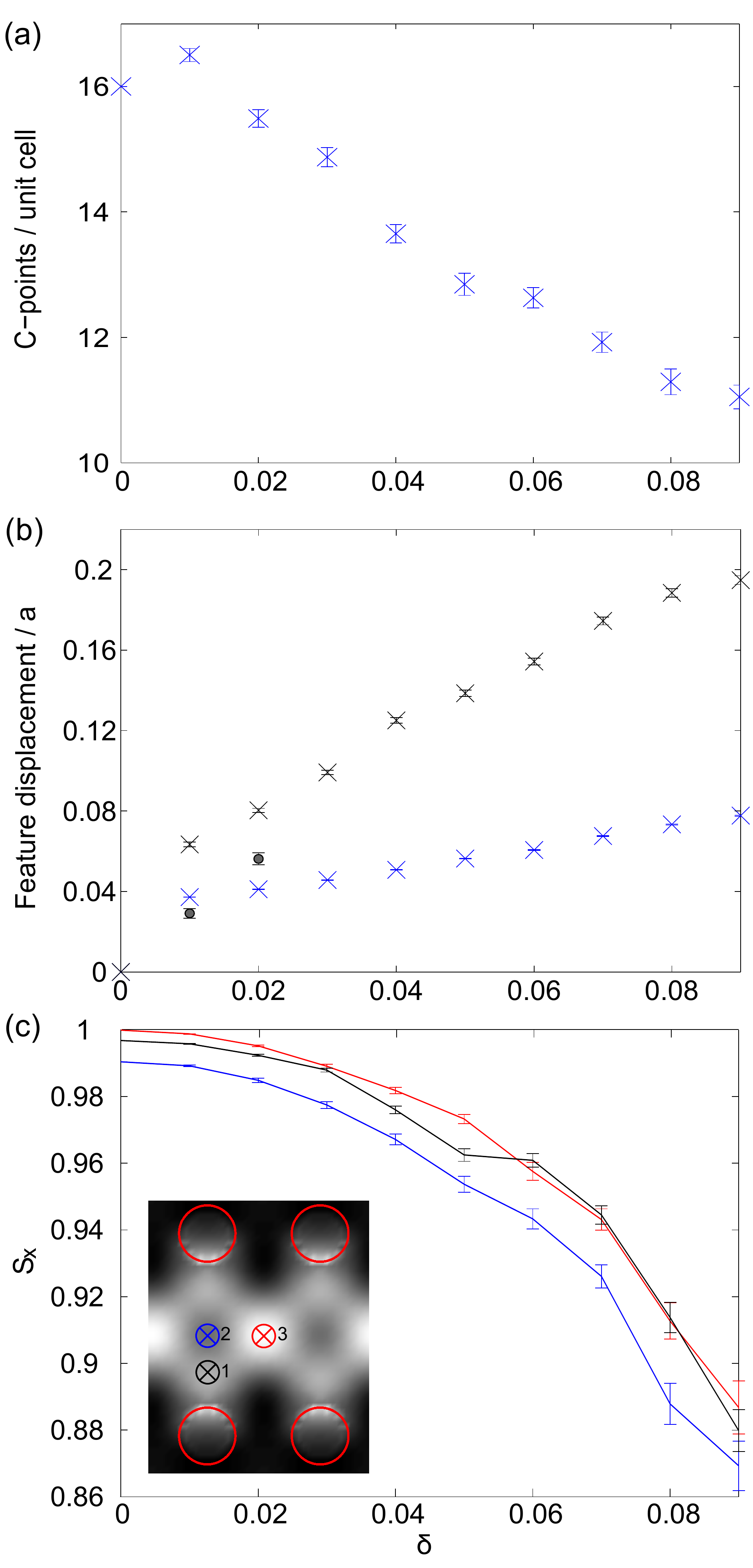}
\caption{(a) The occurrence of C-points as a function of disorder. (b) Crosses (stars): mean distance of C-points (L-line vertices) in simulations of disordered structures from the nearest C-point (L-line) in a simulation of the ideal structure. Circles: mean displacement of C-points when newly created C-points are ignored. (c) The change in polarisation with disorder is plotted for three points, one of right circular polarisation (1) and two of transverse linear polarisation (2,3), for each the relevant Stokes parameter is plotted (3,1,1 respectively). Note that they do not begin at exactly 1; this is a result of finite resolution. The inset shows the locations of points 1,2 and 3 relative to the air holes.}
\label{line}
\end{figure}

Firstly, it is clear that the existence of polarisation singularities is largely robust to the addition of disorder.  A few C-points are eliminated, but the majority survive in displaced locations (see below). For the smallest levels of disorder, the C-points and L-lines cluster very closely to their positions in the ordered waveguide, but they gradually move away with increasing disorder. For $\delta = 0.01$, the C-points move a mean distance of just $0.06a$ away from their positions in the ideal case, and the L-line just $0.04a$, although these distances steadily increase with disorder, as shown in \fig{line}\,(b).

As well as the elimination and displacement of C-points, the addition of disorder also causes their creation.  Such C-points are apparent in \fig{scatter}, where clusters of C-points form that are not associated with any from the ideal waveguide (bottom- and top-left of each panel).  In the ideal structure near these locations, there exist regions where the the ellipticity is high ($>0.9$), despite a C-point not being formed.  The zero-contours of $S_1$ and $S_2$ come very close to crossing in the ideal waveguide (indicated by the arrow in \fig{disorder}\,(b)), and when disorder is introduced, these contours move and can sometimes cross and thus form extra C-points.  In fact, for the smallest levels of disorder considered ($\delta <0.02$), there are on average more C-points found in the disordered waveguides than in the ideal one.  However, for $\delta >0.02$, the number of C-points found per unit cell falls slowly with increasing disorder, until for the highest levels of disorder considered, a mean of 11.1 {\highlight $\pm 0.3$} from the original 16 per unit cell remain. \Fig{line}\,(a) shows a plot of the mean number of C-points found per unit cell as a function of the disorder parameter $\delta$.

The creation of these new C-points increases the average separation between the C-points in disordered structures and those in the ideal structure. The two circles in \fig{line}\,(b) show the mean displacement of C-points when the new C-points are excluded.

To further examine the effects of disorder on the polarisation singularities, we plot the mean values of the Stokes parameters at several key positions in the waveguide. These positions are at a C-point and on an L-line, and are shown in detail by the inset in \fig{line}\,(c). The first point (labelled "1") is located at a C-point in the ideal structure, near the centre of the waveguide. \Fig{line}\,(c) shows the mean value of $S_3$ averaged over all unit cells in all simulations as a function of the disorder parameter $\delta$ at this position.  Being at a (right-handed) C-point means that the value in the ideal waveguide is $S_3 =1$, but this reduces as disorder is introduced and the C-point is displaced or destroyed.

Also shown are two points (label "2" and "3") that are both located on an L-line in the centre of the waveguide.  Both these points have $S_1 = 1$ and $S_3 =0$ in the ideal waveguide, and the mean value of $S_1$ averaged over the disordered structures are shown in \fig{line}\,(c) as a function of $\delta$.  Again, as disorder is introduced, the mean value of $S_1$ deceases, slowly at first, but then faster for higher disorders.  The pattern is very similar to that of the value of the ellipticity $S_3$ at the C-point.  Using the value of $S_1$ as a proxy for the "amount of L-line-ness" at the points considered, it can thus be said that the L-lines and C-points display remarkably similar robustness to the introduction of disorder. 

Taking \fig{scatter} and \ref{line} as a whole, we conclude that the existence of C-points and L-lines are remarkably robust to disorder in the waveguide, and that their positions in the disordered waveguides are on average very close to the positions in the ideal waveguide.

{\highlight Our analysis so far has concentrated on the extremal points of polarisation (i.e. the C-points and L-lines).  However, many applications may only require a large degree of polarisation to be viable. For example, while C-points with $S_3=\pm 1$ are strictly the only locations in the waveguide where a circular dipole will emit light in only one direction (unit directionality), regions possessing a high value of ellipticity $|S_3|$ will emit predominantly in one direction, as the directionality is equal to the ellipticity.  \Fig{line}\,(c) shows the decay of the ellipticity or directionality at the position of the C-point nearest to the centre of the waveguide, and from it we note that disorder parameters of $\delta \le 0.085$ possess a mean directionality of greater than 90\% at this position, and $\delta \le 0.065$ a mean directionality of greater than 95\%.  }

\begin{figure}[h!]
\includegraphics[scale=0.5]{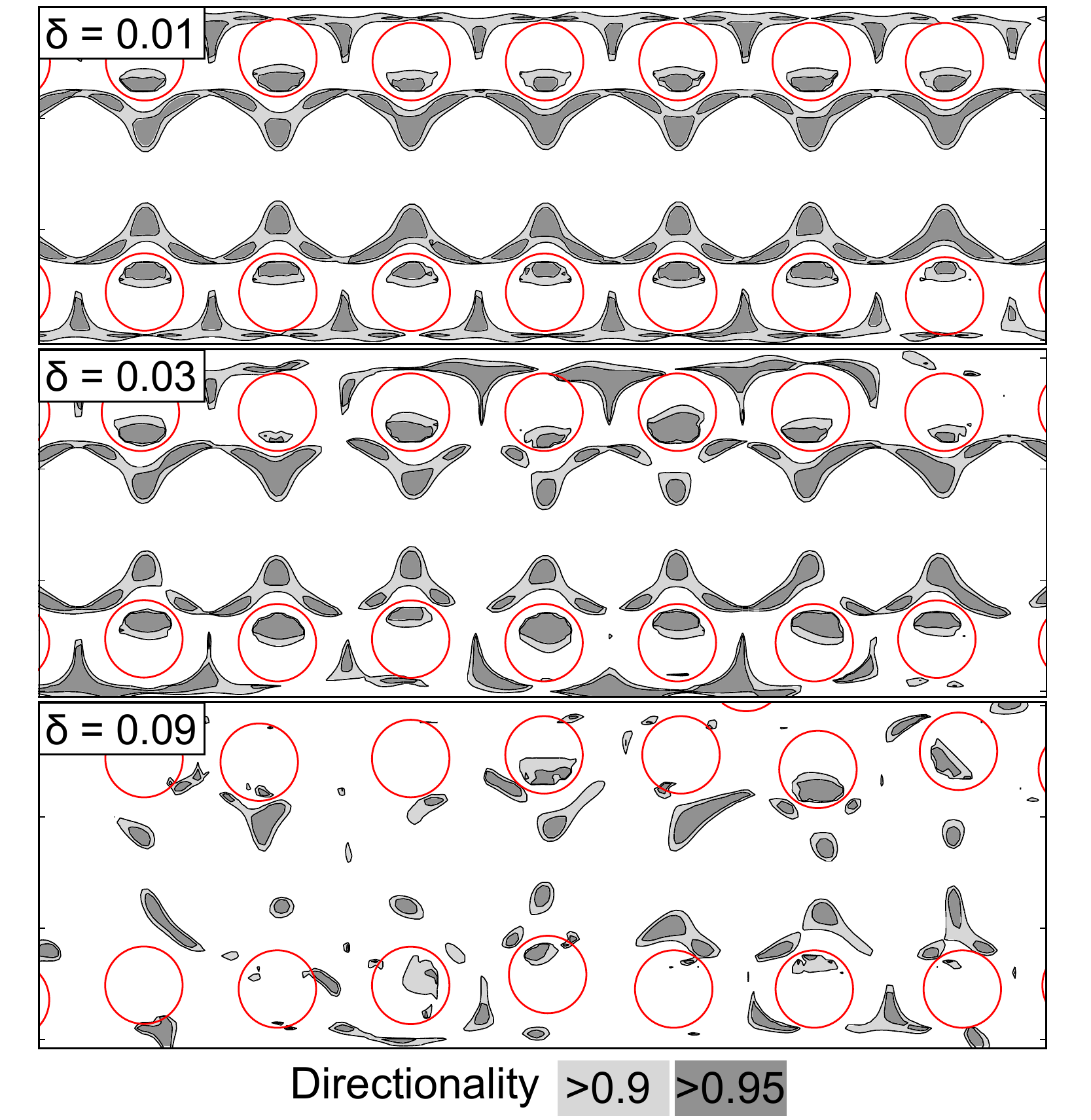}
\caption{Regions of high directionality ($\equiv$ ellipticity) for three specific instances of disorder. Circles show the locations of the air holes.}
\label{elipse}
\end{figure}

{\highlight \Fig{elipse} shows the regions where a circular dipole emitter embedded in the waveguide would possess a high directionality for a particular instance of a disordered waveguide; directionality above 0.9 in the lightly shaded regions and above 0.95 in the darker ones. It can be seen that moderate disorder ($ \delta \le 0.03$, middle panel) does not greatly impact these regions, but that very high disorder ($ \delta = 0.09$, bottom panel) has a large detrimental impact.}

\section{Conclusion}

We have shown that polarisation singularities in photonic crystal waveguides are remarkably robust to disorder in the waveguide, {\highlight at least for this choice of mode}. Using comparisons between experiment and calculations using a similar disorder model to ours in ref. \cite{lodahl2}, we estimate that real photonic crystal waveguides display a level of disorder equivalent to $\delta = 0.003$, more than 3 times smaller than the lowest disorder considered in this paper (note that the comparison used a InGaAsP photonic crystal with a hole spacing of $a=400$\,nm for wavelengths around 1550\,nm). {\highlight Waveguide modes with smaller group velocities may be less robust than the mode studied here (with $v_g=c/7$), but our calculations suggest that even in the case of modes that are an order of magnitude more sensitive to disorder there will still be usable polarisation singularities.} One proposal is to place a quantum dot at a C-point, where the electron spin will be coupled with the path information in the waveguide, useful in quantum information applications.  {\highlight For a quantum emitter placed at the C-point closest to the waveguide core, we find that the emission directionality is 99.6\% $\pm$ 0.3\% for a disorder of $\delta = 0.01$.  Given that site-control placement of quantum dots can be achieved with 50nm accuracy \cite{site_control}, our results indicate that a fabrication yield of 54\% can be expected for emission directionality $>90$\%, a number almost independent of the disorder parameter (for small $\delta$). For comparison of quantum dots grown randomly in the waveguide vicinity 21\% can be expected to have an emission directionality $>90$\% (dropping only to 20\% for small $\delta$). The route to higher yields is to improve the quantum dot placement rather than the disorder in the PhCWG.}

{\highlight By outlining the tolerances to disorder, our results are an important step to realising spin-to-path behaviour in PhCWGs using current fabrication technologies.}

\begin{acknowledgments}

This work has been funded by the project SPANGL4Q, under FET-Open grant number: FP7-284743. RO was sponsored by the EPSRC under grant no. EP/G004366/1, and JGR is sponsored under ERC Grant No. 247462 QUOWSS. DMB acknowledges support from a Marie Curie individual fellowship QUIPS.
This work was carried out using the computational facilities of the Advanced Computing Research Centre, University of Bristol http://www.bris.ac.uk/acrc/.

\end{acknowledgments}

\bibliography{bibliogrpahy}

\newpage


\section*{Supplementary material}

# The stability of polarisation singularities in disordered photonic crystal waveguides: Supplementary Information

Ben Lang,[1] Daryl M. Beggs,[2] Andrew B. Young,[1] John G. Rarity,[1] and Ruth Oulton[1,2]

[1]*Department of Electrical and Electronic Engineering, University of Bristol,*
*Merchant Venturers Building, Woodland Road, Bristol, BS8 1UB, UK*
[2]*Centre for Quantum Photonics, H.H. Wills Physics Laboratory,*
*University of Bristol, Tyndall Avenue, Bristol, BS8 1TL, UK*

# I. CORRELATION ANALYSIS

The supercell length we used for calculations of disordered waveguides was 15 unit cells long, chosen after careful consideration. The supercell should not be too long, in order to conserve computational resources and time, but it must be chosen larger than the correlation length of the disorder. This is the length over which the disorder in one unit cell can still significantly affect the electromagnetic fields in another unit cell.

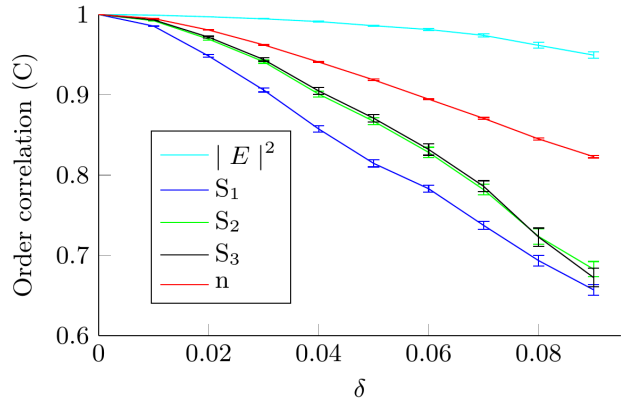

FIG. 1: The order correlation of the Stokes parameters in-plane as a function of disorder. The dielectric function ($n$) order correlation is also plotted.

Figure 1 shows how well the fields in a disordered structure are correlated to the fields in the ideal structure. We call this the "order correlation," and we use it as a measure of how similar the map of each Stokes parameter in the disordered structure are to its equivalent in the ideal counterpart [1]. The order correlation $-1 \leq C \leq 1$ has a value of one when the Stokes parameter map found in the disordered waveguides are identical to those found in the ideal case (to within a constant multiplicative factor) and a value of zero when these fields have no correlation or relation to each other [2].

In fig.1, we plot the order correlation as a function of the disorder parameter $\delta$ for the four Stokes parameters $S_{0,1,2,3}$ and the refractive index profile $n$. The electric field power $|\mathbf{E}|^2$ is only weakly sensitive to the introduction of disorder, as the power density remains predominately in the waveguide core, even for relatively high levels of disorder. For the highest levels of disorder considered here, the power in the disordered structure is still 95% similar to the ideal structure. (Even when the disorder is high enough to close the photonic band-gap entirely, power can still be guided in the waveguide core thanks to the refractive index contrast between the core and the holey region.) In fact, this waveguiding effect is so strong that the order correlation for the electric field power $S_0$ is more robust to disorder than the refractive index profile $n$, which is plotted as a comparison. Note that this means that the spatial profile of the electric field density is similar in a disordered structure to the ideal structure, but this fact can hide real differences in the polarisation structure. The order correlation of the three polarisation parameters $S_{1,2,3}$ appear to be approximately equally sensitive as each other to disorder, as the order correlations drop at similar rates when disorder is introduced, and all three are more sensitive than the power $S_0$. This makes sense physically, as the local polarisation can rotate around the Poincaré sphere, changing all values of $S_{1,2,3}$ whilst simultaneously maintaining the same power $S_0$ as in the ideal waveguide.

Figure 2 shows the mean absolute difference in order correlation between pairs of unit cells in the same simulation as a function of how far apart the pairs are separated. The fields in each of the two unit cells are compared to the fields in the ideal waveguide and the order correlation, $C_i$ and $C_{i+m}$, is calculated as described above. The number $m$ is how far the two unit cells are separated, so for nearest neighbours unit cell, $m = 1$, and in our 15-unit cell long

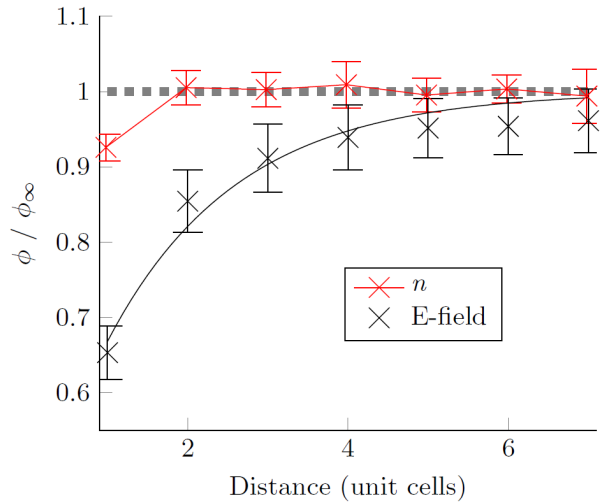

FIG. 2: Normalised $\phi/\phi_\infty$ as a function of distance between unit cells. A value of $\phi/\phi_\infty = 1$ indicates that cells at that distance are causally unrelated, while values $\phi/\phi_\infty < 1$ show that cells still bear a dependence on one another. The refractive index $n$ and an average of $S_{0-3}$ (E-field) are plotted. This is an average over all amounts of disorder.

calculations, $m = 7$ is the maximal separation between pairs of unit cells. The mean absolute difference in order correlation is calculated as $\phi_m = \sum_i^I |C_i - C_{i+m}|/I$ for all pairs of unit cells separated by a distance of $m$ unit cells. Figure 2 plots the value $\phi_m/\phi_\infty$, where $\phi_\infty$ is calculated using unit cells from all other separate calculations (and so guaranteed not to be correlated). A value of $\phi_m/\phi_\infty = 1$ indicates that the pair of unit cells are on average no more correlated to each other than a pair infinitely far apart.

Firstly, notice the correlation of the index profile. Nearest neighbour pairs of unit-cells are somewhat correlated, as they share holes, but those pairs separated by 2 or more unit-cells are no more correlated than those at infinity. This behaviour is as expected, and acts as a check on the method and a useful comparison when using the same measure for the four Stokes parameters. The Stokes parameters all show identical behaviour, the line shown is an average of all of them. Nearest neighbour pairs of unit cells are somewhat related ($\phi_1/\phi_\infty \approx 2/3$), but as the distance increases, the average correlation drops and the value of $\phi_m/\phi_\infty$ approaches 1. The distance at which this happens is a measure of the "disorder correlation length" — i.e. the distance over which disorder in one unit cell can significantly effect the electromagnetic fields in another unit cell. Unit cells within one disorder correlation length cannot be said to be statistically independent of each other, but those far beyond the disorder correlation length are as independent as points at infinity. Ideally the simulation length should be greater than this disorder correlation length, to prevent artefacts arising from the finite length of the waveguide. We find that cells separated by four or more cells are largely independent.

## II. DEPENDENCE ON GROUP INDEX

In this work, we considered a waveguide mode with a group velocity of $v_g = c/7$, chosen for it's significant number of well separated C-points in the waveguide vicinity. However, the dispersion of the photonic crystal waveguide modes means that a wide choice of modes with different group velocities is available, simply by changing the operating wavelength. Slow-light modes with low group velocity are particularly interesting, as they are well known to increase light-matter interactions. In particular, the high density of states afforded by the low group velocity enhances the emission rate of quantum emitters, and the emission efficiency or proportion of photons emitted into the desired

waveguide modes (the "beta-factor"). As these properties are desirable for quantum information applications, it is a reasonable question as to how the results presented in the main paper extend into the slow-light regime. A full analysis of the role of group velocity on the stability of polarisation singularities is beyond the scope of this work, but our initial calculations suggest that C-points persist in waveguide modes of low group velocity and high disorder.

Figure 3 shows the result of calculations of disordered waveguides with group velocities of $v_g \approx c/20$ and $v_g \approx c/40$. The left panels show the zero contours of the Stoke's parameters for the ideal structure ($\delta = 0$), and the right panels for very high disorder of $\delta = 0.09$; far more disorder than could realistically be expected in good samples. Comparison of fig.3 for $\delta = 0$ with fig.2b in the main text is instructive.

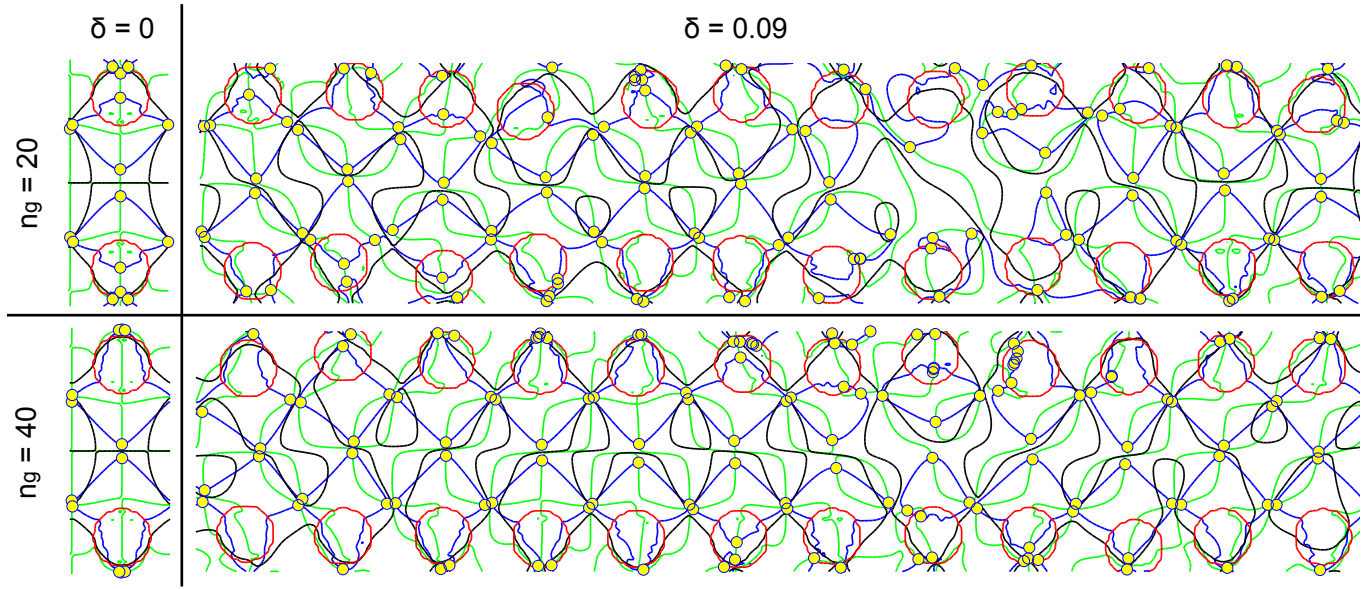

FIG. 3: The zero-crossings of the Stokes parameters ($S_{1;2;3}$ shown by blue, green and black lines respectively) for ideal unit cells and highly disordered ($\delta = 0.09$) supercells. The top and bottom panels show the results for modes with group indices of 20 and 40 respectively. Yellow Circles : C-points, red lines: air holes.

C-points always occur in pairs: one being left-handed ($S_3 = -1$) and the other right ($S_3 = +1$), and the are always separated by at least one L-line ($S_3 = 0$). When the light is slowed, we noticed that the closed contours of $S_1 = 0$ and $S_2 = 0$ become inflated (i.e. the area that they enclose becomes larger), whilst the $S_2 = 0$ lines remain largely unaffected. Thus, although they occur at different frequencies, in some sense we can consider them the same C-points, just shifted in their position as a function of the group velocity. As the group velocity slows, so the pairs of C-points approach each other. At $v_g = c/40$, the C-points can be seen to be very close to each other and their accompanying L-line. As the group velocity approaches zero, the C-points of opposite handedness converge and annihilate each other. (This can be understood in terms of the standing wave being left-right symmetric.) This convergence happens on the L-line which separates the C-points, and so we have a point where $S_1 = S_2 = S_3 = 0$, indicating a node in the light field [3] (such points are also known as V-points [4]). Consequently as the C-points converge on the L-lines we find a rapid drop in the strength of the electric field in the area. This means that the tightly clustered C-points visible for the slow light modes will not be very useful for coupling to emitters, as the local density of states will be relatively low (being located close to a node in the electric field strength).

Upon the introduction of disorder ($\delta = 0.09$, right panels in fig.3), the C-points clearly persist. Note that this is a promising indication only: a full statistical analysis of the behaviour of the C-points as a function of the group velocity is beyond the scope of this work, and we present here only a single instance of a disordered waveguide for illustrative purposes.

However, the polarisation singularities are expected to be significantly affected by occurrences of localised modes at lower group velocities. Localised modes are formed when the disorder is large enough such that the probability of photon scattering becomes significant, characterised by the localisation length. For small disorder, such that the localisation length is larger than the waveguide length, the probability of photon scattering is small and photon transport is ballistic [5]. Moving to lower group velocities increases the optical length of the waveguide, such that it can exceed the localisation length, giving a significant probability that photons are scattered multiple times and localised modes are formed. We expect that the occurrence of the localised modes will inhibit the occurrence of C-points where they occur, as (much like the case of a standing wave with $v_g = 0$ ) the localised mode has left-right symmetry. Therefore we expect that the number of C-points will drop faster with increasing disorder when waveguide modes of lower group velocity are examined.

## III. TOPOLOGICAL CHARGE OF C-POINTS

C-points can come in three possible flavours (not to be confused with their two possible helicities). These flavours are stars, lemons and monstars. Stars have a topological charge of $-1/2$, meaning that along a small clockwise path encircling the star the azimuth angle of the polarisation ellipse rotates half a turn anti-clockwise. Lemons and monstars both have a charge of $+1/2$ [3].

The topological charges of the C-points can be identified visually by colouring the areas between the Stokes contours as shown in fig.4. Within the shaded regions bounded by the zero-contours of both $S_1$ and $S_2$ the azimuth angle (major axis) will lie close to the direction indicated to the right hand side of the figure [6]. From this figure we can read out that the C-points lying in the central symmetry plane, and those at the cell's edge have a positive charge (and are thus either lemons or monstars), while the rest have a negative charge (must be stars).

The topological charge can equivalently be determined by the sign of the expression [7]:

$$\gamma = S_{1x}S_{2y} - S_{2x}S_{1y}. \tag{1}$$

Where $S_{nx}$ ($S_{ny}$) is the partial derivative of the $n$th Stokes parameter with respect to $x$ ($y$).

## IV. LOSS

In our simulations a number of sources within the waveguide supercell emit light into the waveguide. After some time the sources are (gradually) switched off and the light in the waveguide is left to "coast" for a while to allow any radiative modes to dissipate. Thus the total power remaining in the waveguide at the end of a simulation will be influenced both by the amount of power initially radiated by our sources, and by the rate of loss from the waveguide mode. The power each source radiates into the waveguide will be proportional to the density of states (DOS) in the vicinity of the source.

When we introduce disorder it will alter both the DOS and the rate of loss from the waveguide.

As we are in the fairly "fast light" regime the transmission efficiency of the waveguide should vary with $\delta$ as [8]:

$$T = T_{ideal} - A\delta^2. \tag{2}$$

Where $T_{ideal}$ is the transmission of an ideal waveguide with no disorder and A is a constant. In fig.5 we fit this loss model to the power remaining in our simulation's at their end's. The goodness of this fit indicates that loss is the dominant mechanism at work and that changes to the DOS are not playing a significant role.

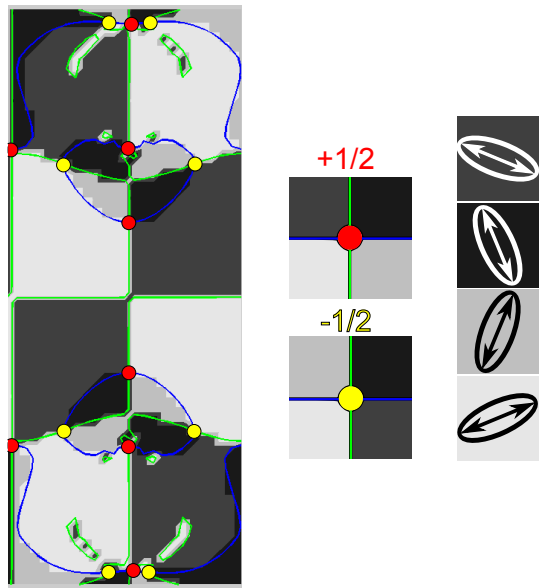

FIG. 4: Finding the charges of C-points. The left panel shows the zero-crossings of $S_{1,2}$ in a single unit cell of waveguide. Here the dominant direction of the major axis of the polarisation ellipse in each region is indicated by the shading (denoted in the key to the left). The charge of a C-point can be read out by looking at the order in which the shades appear in a clockwise circle around the C-point, this is illustrated in the example boxes. C-points of positive (negative) charge are shown in red (yellow).

Localised resonances occur in disordered photonic structures, acting to trap light in a cavity-like way [9]. The occurrence of a localised mode will typically cause a significant change to the DOS [9]. Hence the fit of the loss model to the data in fig.5 suggests that localised modes are not occurring in our calculations.

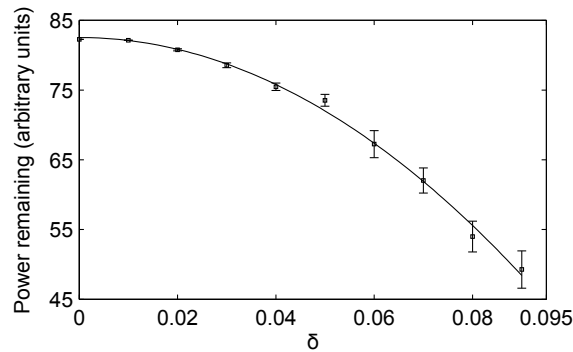

FIG. 5: Points: Power remaining in the central plane of the simulation at it's end, as a function of $\delta$. Line: fit to the data using equation 2.

$$C = \frac{\sum_{\bar{r}}(S(\bar{r}) - \bar{S})(S_{ideal}(\bar{r}) - \bar{S}_{ideal})}{\sqrt{(\sum_{\bar{r}}(S(\bar{r}) - \bar{S})^2)(\sum_{\bar{r}}(S_{ideal}(\bar{r}) - \bar{S}_{ideal})^2)}} \tag{3}$$

 

\end{document}